\documentstyle{article}
\begin{document}
\sloppy
\large
\newcommand{\be}{\begin{equation}}
\newcommand{\ea}{\end{eqnarray}}
\newcommand{\ee}{\end{equation}}
\newcommand{\ba}{\begin{eqnarray}}
\newcommand{\baa}{\begin{eqnarray*}}
\newcommand{\eaa}{\end{eqnarray*}}
\newcommand{\ix}{\int\!\! d^3x}
\newcommand{\iy}{\int\!\! d^3y}
\newcommand{\fba}{\overline{\cal F}^a(x)}
\newcommand{\gba}{\overline{\cal G}^a(x)}
\newcommand{\fa}{{\cal F}^a}
\newcommand{\ga}{{\cal G}^a}
\newcommand{\om}{\widehat\Omega}
\newcommand{\ro}{\widehat\rho}
\newcommand{\ap}{(\Sigma+\epsilon\widetilde\Sigma)}
\newcommand{\bs}{\widehat{B_\Sigma}}
\newcommand{\vu}{\widehat{W}}
\newcommand{\ef}{\widehat{\cal F}}
\newcommand{\gi}{\widehat{\cal G}}
\newcommand{\ach}{\widehat{\cal H}}
\newcommand{\en}{\widehat{\cal N}}

\titlepage

\begin{center}

{\huge Chiral Current Algebras\\
in three--dimensional BF--Theory with boundary}

\vspace{1cm}

{\Large N.Maggiore and P.Provero}

\vspace{1cm}

{\it
Dipartimento di Fisica -- Universit\`a di Genova\\
Istituto Nazionale di Fisica Nucleare -- sez. di Genova\\
Via Dodecaneso, 33 -- 16146 Genova (Italy)}

\end{center}

\vspace{3cm}

\begin{center}
\bf ABSTRACT
\end{center}

{\it We consider the three--dimensional BF--model with planar boundary in the
axial
gauge. We find two--dimensional conserved chiral currents living on the
boundary and satisfying Kac--Moody algebras.}

\vfill
GEF-Th-1/1992 \hfill March 1992
\newpage

\section{Introduction}

In 1981 K.Symanzik faced the problem of studying renormalizable Quantum
Field Theories (QFTs) in presence of a boundary~\cite{sym}. He introduced
boundary conditions for the quantum fields by adding to the action
surface interaction terms, compatible with power counting and locality.

The introduction of a boundary is particularly interesting in topological
QFTs for at least two important reasons.
First, it is known that topological field theories have no local observables
but in the case in which the base manifold has a
boundary~\cite{wit,flope}. Secondly,
all rational conformal field theories
can be classified in terms of Chern-Simons theories built on three-dimensional
manifolds with spatial boundary~\cite{wit,kill}.
This connection is made explicit by noting that
chiral currents satisfying a Kac--Moody algebra live on the two--dimensional
boundary of the three--dimensional base manifold~\cite{wit,dujtru}.

In particular,
the chiral current algebra living on the boundary of a three--dimensional
Chern--Simons theory has been derived in~\cite{blaco} and~\cite{emery}
with an approach   closely
related to Symanzik's ideas and  the existence of chiral currents on the
boundary
and their anomalous Kac--Moody algebra has been derived in the framework of
BRS formalism.
The authors of~\cite{blaco} add to the action
local boundary terms compatible with  power counting, using
a covariant gauge fixing.
The approach followed in~\cite{emery} is different in that the equations of
motion,
rather than the action, are modified by appropriate boundary terms, and
non--covariant axial gauge is preferred.

Recently, many efforts have been made to clarify in general
the relation between
gauge theories in $N$ dimensions and current algebra in $N-1$
dimensions~\cite{flope,dutru2}. The Chern--Simons theory
is not well-suited for this investigation, because of
the difficulties of defining it in an arbitrary number of dimensions, in
particular in the non--abelian case.

This difficulty does not appear in the other important topological QFTs
of the Schwartz--type, the BF models~\cite{report,6132},
which can be defined in any number of dimensions.
The three--dimensional case is of particular physical relevance,
because it coincides with the three--dimensional
Einstein--Hilbert gravity~\cite{report,hoza},
and because
in three dimensions the model can be provided with a true coupling
constant in form of a cosmological constant. The finiteness of this model
has been proved to all
orders in perturbation theory in the Landau gauge~\cite{noi}.

In this paper we study three--dimensional BF theory with a planar boundary
in the axial gauge, following the approach of~\cite{emery}. Such a gauge choice
is quite natural
when studying a QFT with boundary, because the Poincar\'e invariance is lost
{\it a priori} due to the presence itself of the boundary and the main reason
for a covariant choice of the gauge fixing is thus failing. It is on the other
hand
well known that the axial gauge fixing is not a complete one~\cite{piguet}.
It remains indeed
a residual gauge invariance on the planar boundary which can be expressed by
a local Ward identity. The existence of such a ``residual'' Ward identity is
the main advantage of this non--covariant choice, because just from it
one derives the algebra for the conserved chiral currents on the boundary.
Nevertheless, the axial gauge is in general affected by a number of problems
only partially avoided by the ultraviolet finiteness of the theory we consider.
For example we must postulate that the quantum fields vanish at infinity.
The problems deriving from the adoption of an axial gauge in the
three--dimensional Chern--Simons theory with boundary are faced and solved
in~\cite{emery}. The same arguments can be
applied to our case. The aim of this work is rather to investigate the
existence of conserved chiral currents on the boundary and their algebraic
structure.

The plan of the paper is the following: in section 2 we illustrate the model
in unbounded space--time. In section 3 and 4 we find the  free propagators
of the theory with boundary in the ghost and in the gauge sector respectively.
In section 5 we compute the chiral current algebra on the boundary, and
finally in section 6 we draw some concluding remarks.

\section{The model in unbounded flat space--time}

The classical action for the three--dimensional BF--system with cosmological
constant $\lambda\geq 0$ is
\be
S_{BF}=\frac{1}{2}\int_M\!\! d^3x\ \varepsilon^{\mu\nu\rho}\left\{
F^a_{\mu\nu}B^a_\rho+\frac{\lambda}{3}f^{abc}B^a_\mu B^b_\nu B^c_\rho\right\}
\ ,
\ee
where $M$ is the flat space--time, $B_{\mu}^a$ is a one--form and

$$F^a_{\mu\nu} = \partial_\mu A^a_\nu - \partial_\nu A^a_\mu
                + f^{abc} A^b_\mu A^c_\nu \ . $$

As usual, $f^{abc}$ are the structure constants of a compact simple gauge
group $G$.

We make use of the light--cone coordinates
\ba
u &=& x^1 \nonumber\\
z &=& \frac{x^0+x^2}{\sqrt{2}} \\
\bar{z} &=& \frac{x^0-x^2}{\sqrt{2}}\ . \nonumber
\ea
Correspondingly the components of the gauge fields $A_{\mu}^a$ and $B_{\mu}^a$
read
\ba
A_u^a &=& A_1^a \nonumber \\
A^a &=& \frac{A_0^a+A_2^a}{\sqrt{2}} \nonumber \\
\bar{A}^a &=& \frac{A_0^a-A_2^a}{\sqrt{2}} \nonumber \\
B_u^a &=& B_1^a \\
B^a &=& \frac{B_0^a+B_2^a}{\sqrt{2}} \nonumber \\
\bar{B}^a &=& \frac{B_0^a-B_2^a}{\sqrt{2}} \nonumber
\ea
The action for the model becomes
\ba
S_{BF}&=&\int du d^2z \{B^a(\bar{\partial}A_u^a-\partial_u\bar{A}^a+
f^{abc}\bar{A}^b A_u^c)+\bar{B}^a (\partial_uA^a-\partial A_u^a+f^{abc}
A_u^bA^c)\nonumber\\
&&+B_u^a(\partial\bar{A}^a-\bar{\partial}A^a
+f^{abc}A^b\bar{A}^c)+\lambda f^{abc}B^a\bar{B}^bB_u^c\}\ .
\ea
For the aim of this work, a convenient choice of the gauge fixing term is
\ba
S_{gf}=\int du d^2z \{b^aA_u^a+\bar{c}^a(\partial_uc^a+f^{abc}A_u^bc^c+
\lambda f^{abc}B_u^b\phi^c)\nonumber \\
+d^aB_u^a+\bar{\phi}^a(\partial_u\phi^a+f^{abc}A_u^b\phi^c+f^{abc}B_u^b
c^c)\}\ ,
\ea
which corresponds to the ``axial" gauge
\be
A^a_u=B^a_u=0\ .
\ee
In (2.5) $c^a$, $\bar{c}^a$, $b^a$ and $\phi^a$, $\bar{\phi}^a$, $d^a$ are
respectively ghost, antighost and Lagrange multipliers fields for the gauge
fields $A_{\mu}^a$ and $B_{\mu}^a$. In Table 1 are displaied for both fields
and coordinates the canonical dimensions as well as the ghost number and
helicity assignments.
\begin{center}

\begin{tabular}{|c|r|r|r|r|r|r|r|r|r|r|r|r|r|r|r|} \hline
&$z$&$\bar{z}$&$u$
&$A^a$&$\bar{A}^a$&$A^a_u$
&$c^a$&$\bar{c}^a$&$b^a$
&$B^a$&$\bar{B}^a$&$B^a_u$
&$\phi^a$&$\bar\phi^a$
&$d^a$\\ \hline
dim& $-1$    &$-1$  &$-1$
& 1 & 1     &  1
&   1      & 1 &2
&1&1&1
&1&1&2 \\ \hline
helicity&
$-1$&1&0&
1&$-1$&0&
0&0&0&
1&$-1$&0&
0&0&0 \\ \hline
$\Phi\Pi$   &
0&0&0&
0&0&0&
1&$-1$&0&
0&0&0&
1&$-1$&0  \\ \hline
\end{tabular}

\vspace{.3cm}
{\footnotesize
{\bf Table 1.} Dimensions,  $\Phi\Pi$ charges and helicities.}
\end{center}

With the addition to the action of a gauge--fixing term, the general
covariance of the theory, if any, is lost. Here, moreover, the
use of a non--covariant gauge fixing breaks the three--dimensional Poincar\'e
invariance to the two--dimensional one in the plane $\{z,\bar{z}\}$. The gauge
is not completely fixed by (2.5) and, as in~\cite{emery}, the Ward identity
expressing
the residual gauge invariance of the theory will play a key role in determining
the chiral current algebra on the plane boundary $u=0$.

The classical action $S=S_{BF}+S_{gf}$ is invariant under the nilpotent
BRS transformations~\cite{noi}
\ba
sA^a_\mu &=& -(D_\mu c)^a-\lambda f^{abc}B^b_\mu\phi^c \nonumber\\
sc^a     &=& \frac{1}{2}f^{abc}\left (c^bc^c+\lambda \phi^b\phi^c\right)
\nonumber\\
s\bar{c}^a &=& b^a  \nonumber\\
sb^a&=&0 \\
sB^a_\mu &=& -(D_\mu \phi)^a - f^{abc}B^b_\mu c^c\nonumber\\
s\phi^a  &=&f^{abc}\phi^bc^c\nonumber\\
s\bar\phi^a&=&d^a\nonumber\\
sd^a&=&0\ ,\nonumber
\ea
where the covariant derivative is defined by $(D_{\mu}X)^a=\partial_{\mu}
X^a+f^{abc}A_{\mu}^bX^c$. Besides the BRS symmetry (2.7), the theory is
invariant
by the ``parity" coordinate transformations $z\leftrightarrow\bar{z}$,
$u\rightarrow -u$, to which corresponds the discrete field symmetry
\ba
A^a\leftrightarrow\bar{A}^a & \ & B^a\leftrightarrow\bar{B}^a \nonumber \\
A^a_u\rightarrow -A_u^a & \ & B^a_u\rightarrow -B^a_u \nonumber \\
c^a\rightarrow\bar{\phi}^a & \ & \phi^a\rightarrow\bar{c}^a \\
\bar{c}^a\rightarrow-\phi^a & \ & \bar{\phi}^a\rightarrow -c^a \nonumber \\
b^a\rightarrow -b^a & \ & d^a\rightarrow -d^a\nonumber
\ea

At tree level, the generating functional $Z_c(J_{\psi})$ of the connected
Green functions is obtained from the classical action $S(\psi)$ by a Legendre
transformation:
\be
Z_c(J_\psi)=S(\psi)+\int dud^2z \sum_\psi J^a_\psi \psi^a  ,
\ee
where $J_{\psi}^a$ are the sources for the fields, denoted collectively by
$\psi^a$. The following equations of motion for the gauge fields and
multipliers are derived:
\ba
&&
\bar{\partial}B^a_u-\partial_u\bar{B}^a+f^{abc}\bar{A}^bB^c_u
-f^{abc}A^b_u\bar{B}^c+J^a_A=0\nonumber \\&&
\partial_uB^a-\partial B^a_u+f^{abc}A^b_uB^c
-f^{abc}A^bB^c_u+J^a_{\bar{A}}=0\nonumber \\&&
\partial\bar{B}^a-\bar{\partial}B^a+f^{abc}A^b\bar{B}^c-f^{abc}\bar{A}^bB^c
+b^a-f^{abc}\bar{c}^bc^c-f^{abc}\bar\phi^b\phi^c+J^a_{A_u}=0\nonumber \\&&
A^a_u+J^a_b=0 \\&&
\bar{\partial}A^a_u-\partial_u\bar{A}^a+f^{abc}\bar{A}^bA^c_u
+\lambda f^{abc}\bar{B}^bB^c_u+J^a_B=0\nonumber \\&&
\partial_u A^a-\partial A^a_u+f^{abc}A^b_uA^c
-\lambda f^{abc}B^bB^c_u+J^a_{\bar{B}}=0\nonumber \\&&
\partial\bar{A}^a-\bar{\partial}A^a+f^{abc}A^b\bar{A}^c+\lambda
f^{abc}B^b\bar{B}^c
+d^a-f^{abc}\bar{\phi}^bc^c-\lambda f^{abc}\bar{c}^b\phi^c+J^a_{B_u}=0\nonumber
\\&&
B^a_u+J^a_d=0 \nonumber\ ,
\ea
while for the ghost fields one has:
\ba
&&
\partial_u\bar{c}^a+f^{abc}A^b_u\bar{c}^c+f^{abc}B^b_u\bar\phi^c-J^a_c=0
\nonumber \\&&
\partial_uc^a+f^{abc}A^b_uc^c+\lambda f^{abc}B^b_u\phi^c-J^a_{\bar{c}}=0
\nonumber \\&&
\partial_u\bar{\phi}^a+f^{abc}A^b_u\bar{\phi}^c+\lambda f^{abc}B^b_u\bar{c}^c
-J^a_\phi=0 \\&&
\partial_u\phi^a+f^{abc}A^b_u\phi^c+f^{abc}B^b_uc^c-J^a_{\bar\phi}=0\ .
\nonumber
\ea

In order to write a Slavnov identity, we couple external sources to the non
linear variations of quantum fields in (2.7)
\be
S_s=\int dud^2z\left\{\Omega^{a\mu}sA^a_\mu+L^asc^a+\rho^{a\mu}sB^a_\mu
+D^as\phi^a\right\}.
\ee
The dimensions and the Faddeev--Popov charges of the four sources are listed
in Table 2.
\begin{center}

\begin{tabular}{|c|r|r|r|r|} \hline
          &$\Omega^{a\mu}$&$L^a$&$\rho^{a\mu}$&$D^a$\\ \hline
dim&  2    & 3 &  2      & 3 \\ \hline
$\Phi\Pi$   &  $-1$    & $-2$ &  $-1$     & $-2$\\ \hline
\end{tabular}

\vspace{.3cm}
{\footnotesize
{\bf Table 2.} Dimensions and $\Phi\Pi$-charges of the external fields.}
\end{center}

The complete action $\Sigma=S_{BF}+S_{gf}+S_s$ satisfies the Slavnov identity
\be
{\cal S}(\Sigma) =
\int dud^2z\left(
\frac{\delta\Sigma}{\delta\Omega^{a\mu}}\frac{\delta\Sigma}{\delta A^a_\mu} +
\frac{\delta\Sigma}{\delta L^a}\frac{\delta\Sigma}{\delta c^a} +
\frac{\delta\Sigma}{\delta\rho^{a\mu}}\frac{\delta\Sigma}{\delta B^a_\mu}
+\frac{\delta\Sigma}{\delta D^a}\frac{\delta\Sigma}{\delta\phi^a} +
b^a\frac{\delta\Sigma}{\delta\bar{c}^a} +
d^a\frac{\delta\Sigma}{\delta\bar{\phi}^a}\right)=0\ .
\ee
The three--dimensional BF--system exhibits two ghost equations of
motion~\cite{noi}, which
in the axial gauge are local symmetries of the action, broken at the classical
level by terms linear in the quantum fields:
\ba
{\cal F}^a(x)\Sigma &=& \Delta^a_{(f)}(x)\\
{\cal G}^a(x)\Sigma &=& \Delta^a_{(g)}(x)\ ,
\ea
where
\ba
{\cal F}^a(x)&=&\frac{\delta}{\delta c^a}+
f^{abc}\bar{c}^b\frac{\delta}{\delta b^c}+
f^{abc}\bar\phi^b\frac{\delta}{\delta d^c}\\
{\cal G}^a(x)&=&\frac{\delta}{\delta \phi^a}+
f^{abc}\bar\phi^b\frac{\delta}{\delta b^c}
+\lambda  f^{abc}\bar{c}^b\frac{\delta}{\delta d^c}\\
\Delta^a_{(f)}(x)&=&-\partial_\mu\Omega^{a\mu}+\partial_u\bar{c}^a+
f^{abc}(\Omega^{b\mu}A^c_\mu-L^bc^c+\rho^{b\mu}B^c_\mu-
D^b\phi^c)\\
\Delta^a_{(g)}(x)&=&-\partial_\mu\rho^{a\mu}+\partial_u\bar\phi^a+
 f^{abc}(\rho^{b\mu}A^c_\mu-D^bc^c+
\lambda \Omega^{b\mu} B^c_\mu-\lambda  L^b\phi^c)\ .
\ea
Anticommuting (2.14) and (2.15) with the Slavnov operator (2.13) one gets
respectively
\ba
&&
{\cal F}^a(x){\cal S}(\gamma)+B_\gamma({\cal F}^a(x)\gamma-\Delta^a_{(f)}(x)) =
H^a(x)\gamma
\\&&
{\cal G}^a(x){\cal S}(\gamma)+B_\gamma({\cal G}^a(x)\gamma-\Delta^a_{(g)}(x)) =
N^a(x)\gamma\ ,
\ea

where $\gamma$ is a generic functional of the fields, $B_{\gamma}$ is the
anticommuting linearized Slavnov operator
\ba
B_\gamma=&&\!\!\!\!\int dud^2z\left({\ }
\frac{\delta\gamma}{\delta\Omega^{a\mu}}\frac{\delta}{\delta A^a_\mu}
+\frac{\delta\gamma}{\delta A^a_\mu}\frac{\delta}{\delta\Omega^{a\mu}}+
\frac{\delta\gamma}{\delta L^a}\frac{\delta}{\delta c^a}
+\frac{\delta\gamma}{\delta c^a}\frac{\delta}{\delta L^a}\right.\nonumber\\
&&\left.\hspace{.5cm}\mbox{}
+\frac{\delta\gamma}{\delta\rho^{a\mu}}\frac{\delta}{\delta B^a_\mu} +
\frac{\delta\gamma}{\delta B^a_\mu}\frac{\delta}{\delta\rho^{a\mu}}+
\frac{\delta\gamma}{\delta D^a}\frac{\delta}{\delta\phi^a}+
\frac{\delta\gamma}{\delta\phi^a}\frac{\delta}{\delta D^a}\right.\nonumber\\
&&\left.\hspace{.5cm}\mbox{} +
b^a\frac{\delta}{\delta\bar{c}^a} +d^a\frac{\delta}{\delta\bar{\phi}^a}\right)
\ ,
\ea
and
\ba
H^a(x)&=&\partial_{\mu}\frac{\delta}{\delta A_{\mu}^a}-\partial_u b^a
+\sum_{\psi}f^{abc}\psi^b\frac{\delta}{\delta\psi^c}\\
N^a(x)&=&\partial_{\mu}\frac{\delta}{\delta B_{\mu}^a}-\partial_ud^a+
f^{abc}
\left [A_{\mu}^b\frac{\delta}{\delta B_{\mu}^c}+c^b\frac{\delta}
{\delta\phi^c}+d^b\frac{\delta}{\delta b^c}+\bar{\phi}^b\frac{\delta}
{\delta\bar{c}^c}+\rho^{b\mu}\frac{\delta}{\delta\Omega^{c\mu}}+D^b
\frac{\delta}{\delta L^c}\right.\nonumber\\
&&\left.+\lambda \left(B^b_{\mu}\frac{\delta}{\delta A^c_{\mu}}
+\phi^b\frac{\delta}
{\delta c^c}+\bar{c}^b\frac{\delta}{\delta \bar{\phi}^c}+b^b\frac
{\delta}{\delta d^c}+\Omega^{b\mu}\frac{\delta}{\delta\rho^{c\mu}}
+L^b\frac{\delta}{\delta D^c}\right)\right ]
\ea
If, in particular, $\gamma$ stands for the action $\Sigma$, the local operators
$H^a(x)$ and $N^a(x)$ represent exact classical symmetries of the theory
which, written in terms of $Z_c(J_{\psi})$, read
\ba
\lefteqn{
H^a(x)Z_c(J_{\psi})=-\partial_{\mu}J_A^{a\mu}-\partial_u\frac{\delta Z_c}
{\delta J_b^a}+\sum_{\psi}f^{abc}J_{\psi}^b\frac{\delta Z_c}{\delta
J_{\psi}^c}=0}&&\\
\lefteqn{
N^a(x)Z_c(J_{\psi})=-\partial_{\mu}J^{a\mu}_B-\partial_u\frac{\delta Z_c}
{\delta J^a_d}}&&\nonumber\\
&&+f^{abc}\left[J^{b\mu}_B\frac{\delta}{\delta J^{c\mu}_A}
+J^b_{\phi}\frac{\delta}{\delta J^c_c}+J^b_b\frac{\delta}{\delta J^c_d}+
J^b_{\bar{c}}\frac{\delta}{\delta J^c_{\bar{\phi}}}+\rho^{b\mu}
\frac{\delta}{\delta \Omega^{c\mu}}+D^b\frac{\delta}{\delta L^c}\right.
\nonumber\\
&&\left.+\lambda \left(J^{b\mu}_A\frac{\delta}{\delta J^{c\mu}_B}+
J^{b\mu}_c\frac
{\delta}{\delta J^c_{\phi}}+J^b_{\bar{\phi}}\frac{\delta}{\delta
J^c_{\bar{c}}}+J^b_d\frac{\delta}{\delta J^c_b}+\Omega^{b\mu}\frac
{\delta}{\delta \rho^{c\mu}}+L^b\frac{\delta}{\delta D^c}\right)
\right]Z_c=0
\ea
The operators $H^a(x)$ and $N^a(x)$ are the local versions of those found
in~\cite{noi} in the Landau gauge. We have here recovered a general property
of the
axial gauge, $i.e.$  that the Slavnov identities of the theory take the local
form (2.25,26)\cite{piguet}.

\section{Free propagators of the theory with boundary: the ghost sector}

We consider now the model built on the flat space--time ${\bf R}^3$ divided
into two parts ${\bf R}^3_{+}$ and ${\bf R}^3_{-}$ by the plane $u=0$ and we
propose to compute the propagators of the theory
\be
\frac{\delta^2 Z_c(J_{\psi})}{\delta J^a_{\psi_1}(x_1)\delta J^b_{\psi_2}
(x_2)}\Bigr|_{J_{\psi}^a=0}=\Delta^{ab}_{\psi_1\psi_2}(x_1,x_2)=
\langle \psi^a_1(x_1)\psi^b_2(x_2) \rangle\ \ ,
\ee
taking into account the effect of the boundary.

We shall seize upon the procedure illustrated in~\cite{emery}, taking as
outstanding
points of the theory the two requirements of decoupling and locality.

First of all, we demand that the boundary decouples the regions
${\bf R}^3_+$ and ${\bf R}^3_-$ of ${\bf R}^3$. In terms of propagators this
means
\be
\Delta^{ab}_{\psi_1\psi_2}(x_1,x_2)=0\ \ \ \ {\rm if}\ \ u_1u_2<0\ .
\nonumber
\ee
Such a condition is fulfilled by a two--point function of the form
\be
\Delta^{ab}_{\psi_1\psi_2}(x_1,x_2)=\delta^{ab}[\theta_{+}\Delta^{+}_{\psi_1
\psi_2}(x_1,x_2)+\theta_{-}\Delta^{-}_{\psi_1\psi_2}(x_1,x_2)]\ ,
\ee
where $\theta_{\pm}=\theta(\pm u_1)\theta(\pm u_2)$ and $\theta(u)$ is the
step function defined as
\be
\theta(u)=\left\{ \begin{array}{ll}
1 & \mbox{if $u\ge 0$}\\
0 & \mbox{if $u<0$} \end{array} \right.
\ee
We then consider as fundamental the equations of motion of the quantum fields,
modified by boundary terms, in such a way to recover the standard ones (2.10)
and (2.11) away from the boundary (locality condition).

Let us begin analyzing in detail the ghost sector. We then will apply the same
technique to the gauge sector, whose results will be given in the next section.
{}From the free ghost equations of motion
\ba
\partial_u\frac{\delta Z_c}{\delta J^a_{\bar{c}}(x)}-J_c^a(x)=0 &\ &
\partial_u\frac{\delta Z_c}{\delta J^a_c(x)}-J_{\bar{c}}^a(x)=0 \nonumber\\
\partial_u\frac{\delta Z_c}{\delta J^a_{\bar{\phi}}(x)}-J_{\phi}^a(x)=0 &\ &
\partial_u\frac{\delta Z_c}{\delta J^a_{\phi}(x)}
-J_{\bar{\phi}}^a(x)=0
\ea
one gets the equations for the propagators
\be
\begin{array}{lcl}
\partial_{u'}\Delta^{ab}_{c\bar{c}}(x,x')=\delta^{ab}\delta^3(x-x')
&\ \ &
\partial_{u'}\Delta^{ab}_{cc}(x,x')=0\\
\partial_{u'}\Delta^{ab}_{c\bar{\phi}}(x,x')=0&\ \ &
\partial_{u'}\Delta^{ab}_{c\phi}(x,x')=0
\\
\partial_{u'}\Delta^{ab}_{\bar{c}c}(x,x')=\delta^{ab}\delta^3(x-x')&\ \ &
\partial_{u'}\Delta^{ab}_{\bar{c}\phi}(x,x')=0
\\
\partial_{u'}\Delta^{ab}_{\phi\bar{c}}(x,x')=0&\ \ &
\partial_{u'}\Delta^{ab}_{\phi\bar{\phi}}(x,x')=\delta^{ab}\delta^3(x-x')
\\
\partial_{u'}\Delta^{ab}_{\bar{\phi}c}(x,x')=0&\ \ &
\partial_{u'}\Delta^{ab}_{\bar{\phi}\phi}(x,x')=\delta^{ab}\delta^3(x-x')
\end{array}
\ee
The most general solution of equations (3.6) fulfilling the decoupling
condition
and compatible with helicity conservation, scale invariance and regularity,
is
\be
\Delta^{ab}(x,x')=\delta^{ab}[\theta_+\Delta^+
(x,x')+\theta_-\Delta^{-}(x,x')]\ ,
\ee
where
\be
\Delta^{+}(x,x')=\left( \begin{array}{cccc}
0&-T_{\rho}(x,x')&0&\alpha\delta^2(z-z')\\
T_{\rho}(x',x)&0&\beta\delta^2(z-z')&0\\
0&-\beta\delta^2(z-z')&0&-T_{\sigma}(x,x')\\
-\alpha\delta^2(z-z')&0&T_{\sigma}(x',x)&0
\end{array}\right)
\ee
and $\Delta^-(x,x')$ is obtained from $\Delta^+(x,x')$ by parity. In writing
(3.8) we have adopted a matrix notation according to the order $c^a$,
$\bar{c}^a$,
$\phi^a$, $\bar{\phi}^a$. The ghost propagator $\Delta^{ab}(x,x')$ depends
on four constant parameters $\alpha$, $\beta$, $\rho$, $\sigma$ and on the
tempered distribution
\be
T_{\xi}(x,x')\equiv[\theta(u-u')+\xi]\delta^2(z-z')\ .
\ee
The next step is constituted by the most general modification of the free
ghost equations of motion by a boundary term which respects, besides the
locality condition, also the helicity and $\phi\pi$--charge conservation:
\ba
&&\partial_u\bar{c}^a-J_c^a=\delta(u)
[\mu_{+}\bar{c}_{+}^a+\mu_{-}\bar{c}_{-}^a+
k_{+}\bar{\phi}_{+}^a+k_{-}\bar{\phi}_{-}^a]\nonumber\\
&&\partial_u\bar{\phi}^a-J_{\phi}^a=\delta(u)
[\alpha_{+}\bar{c}_{+}^a+\alpha_{-}\bar{c}_{-}^a+
\beta_{+}\bar{\phi}_{+}^a+\beta_{-}\bar{\phi}_{-}^a]\nonumber\\
&&\partial_u\phi^a-J_{\bar{\phi}}^a=-\delta(u)
[\mu_{+}\phi_{-}^a+\mu_{-}\phi_{+}^a+
k_{+}c_{-}^a+k_{-}c_{+}^a]\\
&&\partial_uc^a-J_{\bar{c}}^a=-\delta(u)
[\alpha_{+}\phi_{-}^a+\alpha_{-}\phi_{+}^a+
\beta_{+}c_{-}^a+\beta_{-}c_{+}^a]\nonumber
\ea
where $\mu_{\pm}$, $k_{\pm}$, $\alpha_{\pm}$, $\beta_{\pm}$ are constant
parameters.

In (3.10), the latter two equations are derived from the former two by parity.
The r.h.s. of the above equations depend on the two-dimensional fields which
live on the opposite sides of the boundary:
\be
\psi_{\pm}^a(z,\bar{z})\equiv \lim_{u\rightarrow 0^{\pm}}
\psi^a(u,z,\bar{z})\ .
\ee
Some of the parameters appearing in (3.10) can be eliminated by imposing that
eqs. (3.10) are compatible with each other. Indeed, from the anticommutation
relations between the ghost equations of motion, one finds
\ba
&&k_{+}=k_{-}\equiv k\nonumber\\
&&\beta_{-}=\mu_{+}\nonumber\\
&&\beta_{+}=\mu_{-}\\
&&\alpha_{+}=\alpha_{-}\equiv\nu\nonumber
\ea
{}From the equations (3.10) follow sixteen equations for the free propagators
in presence
of the plane boundary $u=0$, which reduce to eight independent non--linear
equations for the eight parameters $\alpha$, $\beta$, $\rho$, $\sigma$,
$\mu_+$, $\mu_-$, $k$, $\nu$:
\ba
&&(1+\rho)(1-\mu_{+})+\alpha k=0\nonumber\\
&&\sigma(1+\mu_{-})-\alpha k=0\nonumber\\
&&\beta(\mu_{+}-1)+k(1+\sigma)=0\nonumber\\
&&\beta(1+\mu_{-})+ k\rho=0\nonumber\\
&&\rho(1+\mu_{+})+\beta\nu=0\\
&&(1+\sigma)(1-\mu_{-})-\beta\nu=0\nonumber\\
&&\alpha(1+\mu_{+})-\nu\sigma=0\nonumber\\
&&\alpha(1-\mu_{-})+\nu(1+\rho)=0\nonumber
\ea
Of course one could solve directly the non--linear set of equations (3.13),
but the task of finding the solutions of our problem is made much simpler
by the observation that the action $S$ possesses a further discrete simmetry
involving the ghost fields only:
\ba
&&\bar{c}^a\leftrightarrow\phi^a\nonumber\\
&&c^a\leftrightarrow\bar{\phi}^a\ .
\ea
This new symmetry, imposed on the free ghost equations of motion (3.10) and on
the propagators (3.3), respectively gives the following two groups of
constraints
on the parameters:
\ba
&&\mu_{+}=-\mu_{-}\equiv\mu\nonumber\\
&&k=\nu=0
\ea
and
\ba
&&\sigma=-(1+\rho)\nonumber\\
&&\alpha=\beta=0
\ea
Eqs. (3.13) considerably simplifies to
\ba
&&(1+\rho)(1-\mu)=0\nonumber\\
&&\rho(1+\mu)=0
\ea
which has two distinct solutions:

\vspace{1cm}
\begin{center}
\begin{tabular}{|r|r|r|}\hline
&$\rho$&$\mu$\\ \hline
I&0&1\\ \hline
II&-1&-1\\ \hline
\end{tabular}
\end{center}

\vspace{1cm}

Let us take the solution I to discover the corresponding boundary conditions
on the ghost fields. The propagator matrix $\Delta^+(x,x')$ relative to the
half--space ${\bf R}^3_+$ is
\be
\Delta^+(x,x')=\left( \begin{array}{cccc}
0&-\theta(u-u')\delta^2&0&0\\
\theta(u'-u)\delta^2&0&0&0\\
0&0&0&\theta(u'-u)\delta^2\\
0&0&-\theta(u-u')\delta^2&0
\end{array}\right)
\ee
{}From (3.18) we have, for any ghost field $\xi^a(x)$
\ba
&&\lim_{u\rightarrow 0^{+}} \langle c^a(x)\xi^b(x')\rangle =0\nonumber\\
&&\lim_{u\rightarrow 0^{+}} \langle \bar{\phi}^a(x)\xi^b(x')\rangle =0\ .
\ea
We then realize that the ghost field boundary conditions corresponding to the
solution~I are of Dirichlet type
\be
c^a_{+}(z)=\bar{\phi}^a_{+}(z)=0
\ee
and, by parity, one gets from $\Delta^-(x,x')$
\be
c^a_{-}(z)=\bar{\phi}^a_{-}(z)=0\ .
\ee
To conclude the analysis of the first solution, we write the ghost equations
of motion modified by the presence of the plane boundary $u=0$
\ba
&&\partial_u\bar{c}^a+f^{abc}A^b_u\bar{c}^c+f^{abc}B_u^b\bar{\phi}^c
-J_c^a=\delta(u)\bigl(\bar{c}^a_{+}(z)-\bar{c}^a_{-}(z)\bigr)
\nonumber\\
&&\partial_uc^a+f^{abc}A^b_uc^c+\lambda f^{abc}B_u^b\phi^c
-J_{\bar{c}}^a=0
\nonumber\\
&&\partial_u\bar{\phi}^a+f^{abc}A^b_u\bar{\phi}^c+\lambda f^{abc}B_u^b\bar{c}^c
-J_{\phi}^a=0
\\
&&\partial_u\phi^a+f^{abc}A^b_u\phi^c+f^{abc}B_u^bc^c
-J_{\bar{\phi}}^a=\delta(u)\bigl(\phi^a_{+}(z)-\phi^a_{-}(z)\bigr)\ .
\nonumber
\ea
One easily sees that the solution II corresponds to the Dirichlet boundary
conditions
\ba
&&\phi_{+}^a=\phi_{-}^a=0\nonumber\\
&&\bar{c}_{+}^a=\bar{c}_{-}^a=0\ ,
\ea
while the ghost equations of motion read
\ba
&&\partial_u\bar{c}^a+f^{abc}A^b_u\bar{c}^c+f^{abc}B_u^b\bar{\phi}^c
-J_c^a=0
\nonumber\\
&&\partial_uc^a+f^{abc}A^b_uc^c+\lambda f^{abc}B_u^b\phi^c
-J_{\bar{c}}^a=\delta(u)\bigl(c_{+}^a(z)-c_{-}^a(z)\bigr)
\nonumber\\
&&\partial_u\bar{\phi}^a+f^{abc}A^b_u\bar{\phi}^c+\lambda f^{abc}B_u^b\bar{c}^c
-J_{\phi}^a=\delta(u)\bigl(\bar{\phi}_{+}^a(z)-\bar{\phi}_{-}^a(z)\bigr)
\\
&&\partial_u\phi^a+f^{abc}A^b_u\phi^c+f^{abc}B_u^bc^c
-J_{\bar{\phi}}^a=0
\nonumber
\ea

\section{Free propagator of the theory with boundary: the gauge sector}

According to the lines followed in the previous section, from the free
equations of motion of the gauge fields and multipliers follow $8\times 8$
equations for the propagators, many of which are equivalent to each other.
The most general solution compatible with helicity conservation, scale
invariance and regularity is, according to the decoupling condition,
\be
\Delta^{ab}(x,x')=\delta^{ab}[\theta_+\Delta^+(x,x')+\theta_-\Delta^-(x,x')]\ ,
\ee
where we adopted a matrix notation following the order $A$, $\bar{A}$,
$A_u$, $b$, $B$, $\bar{B}$, $B_u$, $d$, and $\Delta^+(x,x')$ is displaied
in table 3, while $\Delta^-(x,x')$ is obtained from $\Delta^+(x,x')$ by a
parity transformation.
The matrix of propagators (4.1) depends on ten constant parameters $a_i$, $i=1,
\dots,10$. The inclusion of a boundary term in the free equations of motion
of the gauge fields and multipliers leads to
\ba
&&
\bar{\partial}B^a_u-\partial_u\bar{B}^a+J^a_A=
\delta(u)[\alpha_1(\bar{A}_{+}^a+\bar{A}_{-}^a)
+\alpha_2\bar{B}_{+}^a+\alpha_3\bar{B}_{-}^a]
\nonumber \\&&
\partial_uB^a-\partial B^a_u+J^a_{\bar{A}}=
\delta(u)[\alpha_1(A_{+}^a+A_{-}^a)
+\alpha_3B_{+}^a+\alpha_2B_{-}^a]
\nonumber \\&&
\partial\bar{B}^a-\bar{\partial}B^a+b^a+J^a_{A_u}=0\nonumber \\&&
A^a_u+J^a_b=0 \\&&
\bar{\partial}A^a_u-\partial_u\bar{A}^a+J^a_B=
\delta(u)[\alpha_3\bar{A}_{+}^a+\alpha_2\bar{A}_{-}^a
+\alpha_4(\bar{B}_{+}^a+\bar{B}_{-}^a)]
\nonumber \\&&
\partial_u A^a-\partial A^a_u+J^a_{\bar{B}}=
\delta(u)[\alpha_2A_{+}^a+\alpha_3A_{-}^a
+\alpha_4(B_{+}^a+B_{-}^a)]
\nonumber \\&&
\partial\bar{A}^a-\bar{\partial}A^a+d^a+J^a_{B_u}=0\nonumber \\&&
B^a_u+J^a_d=0 \nonumber\ ,
\ea
The above equations (4.2), which depend on four constant parameters $\alpha_j$,
$j=1,\dots,4$, respect the parity transformations (2.8) and are compatible
with each other.

{}From the consistency between eqs. (4.2) and the form of the propagator
$\Delta^{ab}(x,x')$ follows a set of non--linear equations for the parameters
$a_i$ and $\alpha_j$
\ba
&&(a_3-a_4)(\alpha_2+1)+a_2\alpha_1=0\nonumber\\
&&(1+a_7)(1-\alpha_3)-a_2\alpha_1=0\nonumber\\
&&(a_6-a_7)(1+\alpha_2)+a_5\alpha_1=0\nonumber\\
&&a_4(1-\alpha_3)-a_1\alpha_1=0\nonumber\\
&&a_9(1+\alpha_2)+a_7\alpha_1=0\nonumber\\
&&a_9(1-\alpha_3)-\alpha_1(1+a_3-a_4)=0\nonumber\\
&&a_{10}(1+\alpha_2)+\alpha_1(a_6-a_7)=0\nonumber\\
&&a_8(1-\alpha_3)-a_4\alpha_1=0\\
&&a_1(1-\alpha_2)-a_4\alpha_4=0\nonumber\\
&&a_5(1+\alpha_3)+\alpha_4(a_6-a_7)=0\nonumber\\
&&a_2(1-\alpha_2)-\alpha_4(1+a_7)=0\nonumber\\
&&a_2(1+\alpha_3)+\alpha_4(a_3-a_4)=0\nonumber\\
&&a_4(1-\alpha_2)-\alpha_4a_8=0\nonumber\\
&&(a_6-a_7)(1+\alpha_3)+\alpha_4a_{10}=0\nonumber\\
&&(1+a_3-a_4)(1-\alpha_2)-\alpha_4a_9=0\nonumber\\
&&a_7(1+\alpha_3)+\alpha_4a_9=0\nonumber
\ea
The equations (4.3) form a set of sixteen
independent non--linear equations for fourteen
parameters.
The introduction of a boundary contribution in the equations of motion~(4.2)
causes a boundary breaking of the Ward identities (2.25) and (2.26), which
we force to be present only at classical level.
We can reach that goal by imposing that the breaking is linear in the quantum
fields\footnote{
We are indebted to O.Piguet for suggesting us this point }:
\ba
\alpha_3&=&\alpha_2\nonumber\\
\alpha_4&=&\lambda\alpha_1.
\ea
The resulting Ward identities are

\ba
\int^{+\infty}_{-\infty}duH^a(x)Z_c(J_{\psi})&=&-\alpha_1(\partial
\bar{A}^a_{+}+\partial\bar{A}^a_{-}+\bar{\partial} A^a_{+}+\bar\partial
A^a_{-})
\nonumber\\
&&-\alpha_2(\partial\bar{B}^a_{+}+\partial\bar{B}^a_{-}+
\bar{\partial}B^a_{+}
+\bar{\partial}B^a_{-})\\
\int^{+\infty}_{-\infty}duN^a(x)Z_c(J_{\psi})&=&-\lambda\alpha_1(\partial
\bar{B}^a_{+}+\partial\bar{B}^a_{-}+\bar{\partial} B^a_{+}+\bar\partial
B^a_{-})
\nonumber\\
&&-\alpha_2(\partial\bar{A}^a_{+}+\partial\bar{A}^a_{-}
+\bar{\partial}A^a_{+}
+\bar{\partial}A^a_{-})
\ea
The identities (4.5) and (4.6) concern the plane $u=0$ and they are
consequences
of the fact that the axial gauge we adopted does not completely fix the gauge.
With all the remarks and prescriptions made in~\cite{emery}, we shall
take~(4.5) and~(4.6) as the Ward identities expressing the residual gauge
invariance of the theory. Equivalently we postulate that
$\int_{-\infty}^{+\infty}du\partial_ub^a(x)
=\int_{-\infty}^{+\infty}du\partial_ud^a(x)=0$. The choice of an axial gauge
in fact does not guarantee that
$b^a(z,\bar{z},\pm\infty)=d^a(z,\bar{z},\pm\infty)=0$.

By imposing the two--dimensional identities (4.5) and (4.6) on the two--point
functions, one gets a further set of constraints for the fourteen parameters
$a_i$ and $\alpha_j$:
\ba
&&\alpha_1a_2+\alpha_2(a_3-a_4)-\alpha_1a_1-\alpha_2a_4=1\nonumber\\
&&\alpha_1a_2+(1+a_7)\alpha_2-a_5\alpha_1-(a_6-a_7)\alpha_2=1\nonumber\\
&&\alpha_1a_7+\alpha_2a_9-\alpha_1a_4-a_8\alpha_2=0\nonumber\\
&&\alpha_1(1+a_3-a_4)+\alpha_2a_9-(a_6-a_7)\alpha_1-a_{10}\alpha_2=0\nonumber\\
&&\alpha_2a_2+\lambda\alpha_1(a_3-a_4)-a_1\alpha_2-\lambda\alpha_1a_4=0\\
&&\alpha_2a_2+\lambda\alpha_1(1+a_7)-a_5\alpha_2-\lambda(a_6-a_7)\alpha_1=0
\nonumber\\
&&\alpha_2a_7+\lambda\alpha_1a_9-\alpha_2a_4-\lambda\alpha_1a_8=1\nonumber\\
&&\alpha_2(1+a_3-a_4)+\lambda\alpha_1a_9-(a_6-a_7)\alpha_2-\lambda\alpha_1a_{10}=1\nonumber
\ea
There are four independent solutions of the two sets of equations (4.3) and
(4.7), which are reported in Table 4.
Notice that the solutions III and IV in the table
depend on the cosmological constant
$\lambda$, and exist only for $\lambda\neq 0$.

\section {Chiral algebra on the plane boundary}

The four solutions in Table 4 can be worked out directly by solving the
equations (4.3) and (4.7). From the matrix of propagators $\Delta^{ab}(x,x')$
we can read the boundary conditions relative to each of them:
\ba
I&:&A^a_{+}=\bar{A}^a_{-}=B^a_{+}=\bar{B}^a_{-}=0\\
II&:&\bar{A}^a_{+}=A^a_{-}=\bar{B}^a_{+}=B^a_{-}=0\\
III&:&A^a_{+}-\sqrt\lambda B^a_{+}=\bar{A}^a_{+}+\sqrt\lambda\bar{B}^a_{+}=
\bar{A}^a_{-}-\sqrt\lambda\bar{B}^a_{-}=  A^a_{-}+\sqrt\lambda B^a_{-}=0\\
IV&:&A^a_{+}+\sqrt\lambda B^a_{+}=\bar{A}^a_{+}-\sqrt\lambda\bar{B}^a_{+}=
\bar{A}^a_{-}+\sqrt\lambda\bar{B}^a_{-}=  A^a_{-}-\sqrt\lambda B^a_{-}=0
\ea
An alternative way to proceed of course could have been the converse one,
{\it i. e.} to figure out all possible behaviours of the gauge fields on
the boundary and then to read from the matrix $\Delta^{ab}(x,x')$ which
conditions on the parameters are derived.

One finds that the conditions written above are the only ones compatible
with the locality and decoupling requirements and, eqs. (5.1,2,3,4) holding,
corresponding to each of them the eqs. (4.3) and (4.7) are easily solved to
give
the four solutions of Table 4.

The solution I corresponds to the modified equations of motion
\ba
&&
\bar{\partial}B^a_u-\partial_u\bar{B}^a+f^{abc}\bar{A}^bB^c_u
-f^{abc}A^b_u\bar{B}^c+J^a_A=-\delta(u)\bar{B}^a_{+}\nonumber \\&&
\partial_uB^a-\partial B^a_u+f^{abc}A^b_uB^c
-f^{abc}A^bB^c_u+J^a_{\bar{A}}=-\delta(u)B^a_{-}\nonumber \\&&
\partial\bar{B}^a-\bar{\partial}B^a+f^{abc}A^b\bar{B}^c-f^{abc}\bar{A}^bB^c
+b^a-f^{abc}\bar{c}^bc^c-f^{abc}\bar\phi^b\phi^c+J^a_{A_u}=0\nonumber \\&&
A^a_u+J^a_b=0 \\&&
\bar{\partial}A^a_u-\partial_u\bar{A}^a+f^{abc}\bar{A}^bA^c_u
+\lambda f^{abc}\bar{B}^bB^c_u+J^a_B=-\delta(u)\bar{A}^a_{+}\nonumber\\&&
\partial_u A^a-\partial A^a_u+f^{abc}A^b_uA^c
-\lambda f^{abc}B^bB^c_u+J^a_{\bar{B}}=-\delta(u)A^a_{-}\nonumber \\&&
\partial\bar{A}^a-\bar{\partial}A^a+f^{abc}A^b\bar{A}^c+\lambda
f^{abc}B^b\bar{B}^c
+d^a-f^{abc}\bar{\phi}^bc^c-\lambda f^{abc}\bar{c}^b\phi^c+J^a_{B_u}=0\nonumber
\\&&
B^a_u+J^a_d=0 \nonumber\ ,
\ea
and the two--dimensional Ward identities (4.5) and (4.6) become
\ba
&&\int^{+\infty}_{-\infty}duH^a(x)Z_c(J_{\psi})=
\partial\bar{B}^a_{+}+\bar{\partial}B^a_{-}\\
&&\int^{+\infty}_{-\infty}duN^a(x)Z_c(J_{\psi})=
\partial\bar{A}^a_{+}+\bar{\partial}A^a_{-}
\ea
{}From (5.6) and (5.7) one gets the following relations between the Green
functions:

\ba
\lefteqn{\bar\partial\langle
B^a(z){\cal A}^{a_1\ldots a_N}_{z_1,\ldots,z_N}
{\cal B}^{a_{N+1}\ldots a_M}_{z_{N+1},\ldots,z_M}
\rangle_{-}=}&&\\
&=&-\delta_{1N}\delta_{NM}\delta^{aa_1}\partial\delta^2(z-z_1)\nonumber\\
&&+\sum_{k=1}^N f^{aa_kb}\delta^2(z-z_k)
\langle
A^b(z)A^{a_1}(z_1)\ldots \widehat{A^{a_k}(z_k)}
\ldots A^{a_N}(z_N){\cal B}^{a_{N+1}\ldots a_M}_{z_{N+1},\ldots,z_M}
\rangle_{-}\nonumber\\
&&+\sum_{k=N+1}^M f^{aa_kb}\delta^2(z-z_k)
\langle
B^b(z){\cal A}^{a_1\ldots a_N}_{z_1,\ldots,z_N}B^{a_{N+1}}(z_{N+1})\ldots
\widehat{B^{a_k}(z_k)}\ldots B^{a_M}(z_M)\rangle_{-}\nonumber\ ,
\ea
\ba
\lefteqn{
\bar\partial\langle
A^a(z)
{\cal A}^{a_1\ldots a_N}_{z_1,\ldots,z_N}
{\cal B}^{a_{N+1}\ldots a_M}_{z_{N+1},\ldots,z_M}
\rangle_{-}= }&&\\
&=&-\delta_{0N}\delta_{1M}\delta^{aa_1}\partial\delta^2(z-z_1)\nonumber\\
&&+\lambda\sum_{k=1}^N f^{aa_kb}\delta^2(z-z_k)
\langle
B^b(z)A^{a_1}(z_1)\ldots \widehat{A^{a_k}(z_k)}
\ldots A^{a_N}(z_N){\cal B}^{a_{N+1}\ldots a_M}_{z_{N+1},\ldots,z_M}
\rangle_{-}\nonumber\\
&&+\sum_{k=N+1}^M f^{aa_kb}\delta^2(z-z_k)
\langle
A^b(z){\cal A}^{a_1\ldots a_N}_{z_1,\ldots,z_N}
B^{a_{N+1}}(z_{N+1})\ldots
\widehat{B^{a_k}(z_k)}\ldots B^{a_M}(z_M)\rangle_{-}\nonumber\ ,
\ea

where $(z)$ stands for $(z,\bar{z})$, the hat means omission of its argument
and
\ba
{\cal A}^{a_1\ldots a_N}_{z_1,\ldots,z_N}&\equiv&
A^{a_1}(z_1)\ldots A^{a_N}(z_N)\\
{\cal B}^{a_{N+1}\ldots a_M}_{z_{N+1},\ldots,z_M}&\equiv&
B^{a_{N+1}}(z_{N+1})\ldots B^{a_M}(z_M)\ .
\ea
Let us consider first the case $\lambda=0$. For the operators
\be
A^a(z,\bar{z})=\lim_{u\rightarrow 0^-}A^a(x)\quad\quad
B^a(z,\bar{z})=\lim_{u\rightarrow 0^-}B^a(x)
\ee
we derive the conservation laws
\be
\bar{\partial}A^a(z,\bar{z})=\bar{\partial}B^a(z,\bar{z})=0
\ee
and the commutation relations
\ba
&&\left[A^a(z),A^b(z')\right]=0\nonumber\\
&&\left[B^a(z),B^b(z')\right]=f^{abc}\delta(z-z')B^c(z)\\
&&\left[A^a(z),B^b(z')\right]=f^{abc}\delta(z-z')A^c(z)
-\delta^{ab}\delta'(z-z')\nonumber
\ea
This algebra can be interpreted as a semidirect sum of the Kac--Moody
algebra satisfied by $B$ and its adjoint representation with a central
extension
in the mixed commutators.

If $\lambda>0$, the operators
\ba
K^a(z,\bar{z})&=&\frac{1}{2}\lim_{u\rightarrow 0^-}
\left(\frac{1}{\sqrt\lambda}A^a(x)+B^a(x)\right)\nonumber\\
T^a(z,\bar{z})&=&\frac{1}{2}\lim_{u\rightarrow 0^-}
\left(-\frac{1}{\sqrt\lambda}A^a(x)+B^a(x)\right)
\ea
are conserved currents:
\be
\bar{\partial}K^a(z,\bar{z})=\bar{\partial}T^a(z,\bar{z})=0
\ee
and satisfy the commutation relations
\ba
&&\left[K^a(z),K^b(z')\right]=f^{abc}\delta(z-z')K^c(z)
-\frac{1}{2\sqrt\lambda}\delta^{ab}\delta'(z-z')\nonumber\\
&&\left[T^a(z),T^b(z')\right]=f^{abc}\delta(z-z')T^c(z)+
\frac{1}{2\sqrt\lambda}\delta^{ab}\delta'(z-z')\\
&&\left[K^a(z),T^b(z')\right]=0\nonumber
\ea
which is the direct sum of two Kac--Moody algebras with central extension
$\pm\frac{1}{2\sqrt\lambda}$.

The case $\lambda\neq 0$ is not a common feature of the generic
$d$--dimensional
BF--system. In four dimensions, for instance, the model cannot be provided
of a cosmological constant. We
could therefore guess that only the algebra corresponding to $\lambda=0$
survives in such cases.

By parity we can find the conservation laws and the algebraic structure
which live on the opposite side of the boundary.

The solution II is easily recognized to give the same algebraic structure
of the first solution, therefore we skip to the solution III, for which
the equations of motion read
\ba
&&
\bar{\partial}B^a_u-\partial_u\bar{B}^a+f^{abc}\bar{A}^bB^c_u
-f^{abc}A^b_u\bar{B}^c+J^a_A=\delta(u)\frac{1}{\sqrt\lambda}
(\bar{A}^a_{+}+\bar{A}^a_{-})
\nonumber \\&&
\partial_uB^a-\partial B^a_u+f^{abc}A^b_uB^c
-f^{abc}A^bB^c_u+J^a_{\bar{A}}=\delta(u)\frac{1}{\sqrt\lambda}
(A^a_{+}+A^a_{-})\nonumber \\&&
\partial\bar{B}^a-\bar{\partial}B^a+f^{abc}A^b\bar{B}^c-f^{abc}\bar{A}^bB^c
+b^a-f^{abc}\bar{c}^bc^c-f^{abc}\bar\phi^b\phi^c+J^a_{A_u}=0\nonumber \\&&
A^a_u+J^a_b=0 \\&&
\bar{\partial}A^a_u-\partial_u\bar{A}^a+f^{abc}\bar{A}^bA^c_u
+\lambda f^{abc}\bar{B}^bB^c_u+J^a_B=\delta(u){\sqrt\lambda}
(\bar{B}^a_{+}+\bar{B}^a_{-})\nonumber\\&&
\partial_u A^a-\partial A^a_u+f^{abc}A^b_uA^c
-\lambda f^{abc}B^bB^c_u+J^a_{\bar{B}}=\delta(u){\sqrt\lambda}
(B^a_{+}+B^a_{-})\nonumber \\&&
\partial\bar{A}^a-\bar{\partial}A^a+f^{abc}A^b\bar{A}^c+\lambda
f^{abc}B^b\bar{B}^c
+d^a-f^{abc}\bar{\phi}^bc^c-\lambda f^{abc}\bar{c}^b\phi^c+J^a_{B_u}=0\nonumber
\\&&
B^a_u+J^a_d=0 \nonumber\ ,
\ea
The Ward identities expressing the residual gauge invariance on the boundary
are
\ba
&&\int^{+\infty}_{-\infty}duH^a(x)Z_c(J_{\psi})=
-\frac{1}{\sqrt\lambda}
(\partial\bar{A}^a_{+}+\partial\bar{A}^a_{-}
+\bar{\partial}A^a_{+}+\bar{\partial}A^a_{-})\\
&&\int^{+\infty}_{-\infty}duN^a(x)Z_c(J_{\psi})=
-(\partial\bar{B}^a_{+}+\partial\bar{B}^a_{-}
+\bar{\partial}B^a_{+}+\bar{\partial}B^a_{-})
\ea
The relations between Green functions are better understood in terms of the
fields
\ba
P^a(x)&=&\frac{A^a(x)+\sqrt\lambda B^a(x)}{\sqrt{1+\lambda}}\\
Q^a(x)&=&\frac{A^a(x)-\sqrt\lambda B^a(x)}{\sqrt{1+\lambda}}\ ,
\ea
for which the Dirichlet boundary conditions hold:
\be
Q^a_+(z,\bar{z})=\bar{P}^a_+(z,\bar{z})=\bar{Q}^a_-(z,\bar{z})=
P^a_-(z,\bar{z})=0\ .
\ee
For the new variables the identities (5.19) and (5.20) imply the conservation
laws
\be
\bar{\partial}P^a_{+}(z,\bar{z})=0\quad\quad\partial\bar{Q}^a_{+}
(z,\bar{z})=0
\ee
and the chiral current  algebra
\ba
\lefteqn{
\bar\partial\langle
P^a(z)
{\cal P}^{a_1\ldots a_N}_{z_1,\ldots,z_N}
\bar{\cal Q}^{a_{N+1}\ldots a_M}_{\bar{z}_{N+1},\ldots,\bar{z}_M}
\rangle_{+}=}\\
&=&c(\lambda)_+
\delta_{1N}\delta_{NM}\delta^{aa_1}\partial\delta^2(z-z_1)+
c(\lambda)_-\delta_{0N}\delta_{1M}\delta^{aa_1}
\bar\partial\delta^2(z-z_1)
\nonumber\\
&&
-c(\lambda)_+\sum_{i=1}^N f^{aa_ib}\delta^2(z-z_i)
\langle
P^b(z)P^{a_1}(z_1)\ldots \widehat{P^{a_i}(z_i)}
\ldots P^{a_N}(z_N)
\bar{\cal Q}^{a_{N+1}\ldots a_M}_{\bar{z}_{N+1},\ldots,\bar{z}_M}
\rangle_{+}\nonumber\\
&&-c(\lambda)_-\sum_{i=N+1}^M f^{aa_ib}\delta^2(z-z_i)
\langle
\bar{Q}^b(z)
{\cal P}^{a_1\ldots a_N}_{z_1,\ldots,z_N}
\bar{Q}^{a_{N+1}}(z_{N+1})\ldots
\widehat{\bar{Q}^{a_i}(z_i)}\ldots \bar{Q}^{a_M}(z_M)\rangle_{+}\nonumber\ ,
\ea
and
\ba
\lefteqn{
\partial\langle
\bar{Q}^a(z)
{\cal P}^{a_1\ldots a_N}_{z_1,\ldots,z_N}
\bar{\cal Q}^{a_{N+1}\ldots a_M}_{\bar{z}_{N+1},\ldots,\bar{z}_M}
\rangle_{+}=}\\
&=&c(\lambda)_-
\delta_{1N}\delta_{NM}\delta^{aa_1}\partial\delta^2(z-z_1)+
c(\lambda)_+ \delta_{0N}\delta_{1M}\delta^{aa_1}
\bar\partial\delta^2(z-z_1)
\nonumber\\
&&-c(\lambda)_-\sum_{i=1}^N f^{aa_ib}\delta^2(z-z_i)
\langle
P^b(z)P^{a_1}(z_1)\ldots \widehat{P^{a_i}(z_i)}
\ldots P^{a_N}(z_N)
\bar{\cal Q}^{a_{N+1}\ldots a_M}_{\bar{z}_{N+1},\ldots,\bar{z}_M}
\rangle_{+}\nonumber\\
&&-c(\lambda)_+\sum_{i=N+1}^M f^{aa_ib}\delta^2(z-z_i)
\langle
\bar{Q}^b(z)
{\cal P}^{a_1\ldots a_N}_{z_1,\ldots,z_N}
\bar{Q}^{a_{N+1}}(z_{N+1})\ldots
\widehat{\bar{Q}^{a_i}(z_i)}\ldots \bar{Q}^{a_M}(z_M)\rangle_{+}\nonumber\ ,
\ea
where
\be
c(\lambda)_\pm=\frac{\sqrt\lambda(1\pm\sqrt\lambda)}{1+\lambda}\
\ee
and
\ba
{\cal P}^{a_1\ldots a_N}_{z_1,\ldots,z_N}&\equiv&
P^{a_1}(z_1)\ldots P^{a_N}(z_N)\\
\bar{\cal Q}^{a_{N+1}\ldots a_M}_{\bar{z}_{N+1},\ldots,\bar{z}_M}&\equiv&
\bar{Q}^{a_{N+1}}(\bar{z}_{N+1})\ldots \bar{Q}^{a_M}(\bar{z}_M)\ .
\ea
It is easily seen that eqs (5.25) and (5.26) are compatible only if
$\lambda=1$.
We finally derive for the operators
\be
K^a(z,\bar{z})=-\sqrt{2}\lim_{u\rightarrow o^+}P^a(x)\ \ \ \
\bar{T}^a(z,\bar{z})=-\sqrt{2}\lim_{u\rightarrow o^+}\bar{Q}^a(x)
\ee
the conservation laws
\be
\bar\partial K^a(z,\bar{z})=\partial \bar{Q}^a(z,\bar{z})=0
\ee
and the chiral current algebra
\ba
&&\left [K^a(z),K^b(z')\right]=
f^{abc}\delta(z-z')K^c(z)+\frac{1}{2}\delta^{ab}\delta'(z-z')\nonumber\\
&&\left [\bar{T}^a ( \bar{z} ) , \bar{T}^b ( \bar{z} ' ) \right
] =
f^{abc}\delta(z-z')\bar{T}^c(\bar{z})-\frac{1}{2}\delta^{ab}
\delta'(\bar{z}-\bar{z} ')\\
&&\left [ K^a({z}),\bar{T}^b(\bar{z} ')\right ]=0\nonumber\ .
\ea
The solution IV corresponds to the exchange $K^a(x)\leftrightarrow T^a(x)$.

\section{Conclusions}

We studied the three--dimensional BF model with a planar boundary in
the axial gauge.
We found the most general equations of motion for the quantum fields
of the theory with boundary compatible with the fundamental requirements of
locality and decoupling.

The choice of an axial gauge allowed us to write the Slavnov
identities expressing the residual gauge invariance on the planar boundary.
{}From these identities we found three possible distinct anomalous chiral
current
algebras, with two conserved chiral currents for each of them.

In the case $\lambda=0$ we obtain a semidirect sum of a Kac--Moody algebra
satisfied  by a chiral conserved current and its adjoint representation acting
on a current of the same chirality (5.14). For a generic $\lambda\neq 0$ we
find
a direct sum of two independent Kac--Moody algebras satisfied by conserved
currents of the same chirality (5.17).
In the particular case $\lambda=1$, we find a third possible algebraic
structure: a direct sum of Kac--Moody algebras satisfied by conserved currents
of opposite chirality (5.32).

\vspace{2cm}

{\bf Acknowledgements}

\vspace{1cm}

We would like to thank A.Blasi R.Collina, S.Emery and O.Piguet for
useful discussions and comments. N.M. would like to express his gratitude to
the D\'epartement de Physique Th\'eorique de l'Universit\'e de Gen\`eve, where
part of this work has been done.

\end{document}